\documentclass[10pt]{article}
\usepackage[utf8]{inputenc}
\usepackage[T1]{fontenc}
\usepackage{amsmath}
\usepackage{amsfonts}
\usepackage{amssymb}
\usepackage[version=4]{mhchem}
\usepackage{stmaryrd}
\usepackage{graphicx}
\usepackage[export]{adjustbox}
\usepackage{enumitem}
\usepackage[round]{natbib}
\usepackage{multirow} 
\usepackage{longtable} 

\graphicspath{ {./images/} }

\title{On Robust Measures of Spatial Correlation}
\author{Vincenzo Nardelli, Giuseppe Arbia}
\date{}
\begin{document}
\maketitle

\begin{abstract}
As a rule statistical measures are often vulnerable to the presence of outliers and spatial correlation coefficients, critical in the assessment of spatial data, remain susceptible to this inherent flaw. In contexts where data originates from a variety of domains (such as, e. g., socio-economic, environmental or epidemiological disciplines) it is quite common to encounter not just anomalous data points, but also non-normal distributions. These irregularities can significantly distort the broader analytical landscape often masking significant spatial attributes. This paper embarks on a mission to enhance the resilience of traditional spatial correlation metrics, specifically the Moran Coefficient (MC), Geary's Contiguity ratio (GC), and the Approximate Profile Likelihood Estimator (APLE) and to propose a series of alternative measures. Drawing inspiration from established analytical paradigms, our research harnesses the power of influence function studies to examine the robustness of the proposed novel measures in the presence of different outlier scenarios.
\end{abstract}

\section{Introduction}
Spatial correlation analysis is an essential tool in spatial data analysis especially when dealing with socio-economic, environmental and epidemiological empirical phenomena. The analysis of this phenomena often results in encountering data that diverge from the traditional normal distribution. This deviation, commonly referred to as non-normality, is typically characterized by features such as skewness and kurtosis, or the emergence from a mixture of other distributions, often due to the presence of outliers. In environmental studies, this is evident in the distribution of pollution across different areas. In socio-economic research, key indicators such as income and population density often show spatial patterns with non-normal distributions. Similarly, in epidemiological studies, the spread of epidemics in space is a clear manifestation of these complex data mixtures. 

A central question faced by researchers is determining whether these deviations from normality are associated to the spatial distribution of observations, potentially influenced by the autocorrelation and the presence of hotspots. However, most statistical measures are very sensitive to the presence of outliers and spatial correlation coefficients are not an exception to this rule. Indeed, spatial outliers are very common in practice when data are observed both on regular lattices (e. g. in satellite images) and in non-lattice data such as administrative partitions like countries or regions. In these cases, the presence of few exceptional observations may dramatically distort the picture and hide interesting spatial features. 

In this paper we introduce methods for robustizing traditional spatial correlation measures, such as the Moran Coefficient (MC; see \citep{moran1948interpretation}, Geary's Contiguity ratio (GC; see \citep{geary1954contiguity} and the Approximate Profile Likelihood Estimator (APLE; see \citep{li2007beyond, li2012one}). Although not in the same way, all three measures are sensitive to observations that may disproportionately influence the measurement of spatial dependence. We propose some robust alternatives to the spatial correlation traditional measures extending to the spatial case the general notion of robust correlation introduced by \cite{gnanadesikan1972robust}. Following the traditional approach in robust statistics, we will base our analysis on the exam of influence function \citep{hampel1974influence} through which we will compare the robustness performances of the traditional measures with those of our proposed alternatives. For the sake of comparisons, we will make use of Monte Carlo experiments with which we simulate sets of spatial data, where we surreptitiously introduce different simulated outlier conditions including extreme cases of skewness and kurtosis. The following Section 2 will concern the presentation of some spatial correlation measures. Sections 3 and 4 will be devoted to the study of influence functions and to robust estimation respectively. Section 5 will contain the conclusions.

\section{Some measures of spatial correlation}
At the heart of all measures of spatial correlation studied in the literature we find the definition of the so-called weight matrix $(W)$ which accomplishes the task of describing the topology of the spatial system on which the data are laid. As we will see later, it also plays a fundamental role in the analysis of the effect of outliers on spatial correlation measures. Suppose we have $n$ observations of a random variable $Z$ say $z=\left(z_{1}, z_{2}, \ldots, z_{n}\right)$, which, without loss of generality, are assumed centered around the mean and distributed on (possibly irregoular) lattice locations. The generic entry $w_{i j} \in W$, expresses the level of connectedness

between location \( i \) and location \( j \), where
\[ w_{ij} = 
\begin{cases} 
0 & \text{if } i=j \\
>0 & \text{if } j \in N(i) \\

0 & \text{otherwise} 
\end{cases}
\]
with $N(i)$ the set of locations connected with location $i$. 

Consequently, $\sum_{j=1}^{n} w_{i j}=\eta_{i}$ is the connectivity of location $i$ (or weighted outdegree in the graph theory terminology. See \cite{bang2007theory}), and $\bar{\eta}=n^{-1} \sum_{i=1}^{n} \eta_{i}$ is the average connectivity of the spatial system. $W$ is often row-standardized so that $\eta_{i}=1$ for each $\mathrm{i}$ and $\bar{\eta}=n$. In the remainder of this paper we will always assume this row standardization to hold for the W matrix. Given these definitions the weighted average of the neighbours of location $i$ :

\begin{equation}
L\left[z_{i}\right]=\sum_{i=1}^{n} w_{i j} z_{j}
\label{eq:lag}
\end{equation}

(with $w_{i j} $ the generic entry of a row-standardized $W$ matrix), assumes the role of the spatially lagged variable by analogy to the time series definition. Generalizing, we have $L[Z]=W Z$.

The Moran coefficient \citep{moran1950notes} can then be defined as:

\begin{equation}
MC=\frac{\sum_{i=1}^{n}\left(z_{i}\right) L\left[z_{i}\right]}{\sum_{i=1}^{n}\left(z_{i}\right)^{2}}=\frac{Z^{T} L(Z)}{Z^{T} Z}
\label{eq:MC}
\end{equation}

Equation \ref{eq:MC} is the ratio between the spatial autocovariance and the variance of $Z$. However, it is improperly referred to as a spatial correlation coefficient. Indeed, it assumes the form of a correlation only if $\operatorname{Var}(z)=\operatorname{Var}(L[z])$, which is not the case unless in trivial situations. As a consequence, its range is narrower than the interval $[-1 ; 1]$ (see \cite{arbia1989statistical}) and it depends on the extreme eigenvalues of $W$ \citep{griffith2010moran}. 

The APLE statistics \citep{calder2007some} was introduced to tackle one important limitation of MC: it is good estimator of the parameter of a spatial autoregressive model \citep{cressie1993aggregation} only in trivial cases of no practical interest. As an alternative, APLE assumes the following form:

\begin{equation}
APLE=\frac{1}{2} \frac{\left[Z^{T} W^{T} Z+Z^{T} W Z\right]}{Z^{T} W^{T} W Z+\operatorname{tr}\left(W^{2}\right) Z^{T} Z / n}=\frac{1}{2} \frac{\left[L(Z)^{T} Z+Z^{T} L(Z)\right]}{L(Z)^{T} L(Z)+\operatorname{tr}\left(W^{2}\right) Z^{T} Z / n}
\label{eq:APLE}
\end{equation}

which, when $W$ is symmetric, boils down to a spatial autocovariance with a different normalizing factor in the denominator with respect to the MC coefficient:

\begin{equation}
APLE=\frac{Z^{T} L(Z)}{L(Z)^{T} L(Z)+\operatorname{tr}\left(W^{2}\right) Z^{T} Z / n}
\label{eq:APLE}
\end{equation}

Finally, Geary's coefficient \citep{geary1954contiguity} is not a correlation measure, and is expressed as the ratio of two sums of squares:

\begin{equation}
GC=\frac{(2 n \bar{\eta})^{-1} \sum_{i=1}^{n} \sum_{j=1}^{n} w_{i j}\left(z_{i}-z_{j}\right)^{2}}{(n-1)^{-1} \sum_{i=1}^{n}(z)^{2}}
\label{eq:GC}
\end{equation}

and, rather counterintuitively, falls in the range $[0 ; 2]$ revealing negative spatial correlation if greater than 1, positive if lower than 1, and no correlation if it is equal to 1. Such a measure can however be easily normalized in the more intuitive interval $[-1 ; 1]$ considering the transform 1 - GC.

Although different if used in a descriptive context, MC and GC are inferentially equivalent.

\section{Influence functions in space}
In general, if we define $\hat{\theta}$ as an estimator of a generic parameter $\theta$, based on $\mathrm{n}$ observations, two finite sample versions of Hampel’s influence function \citep{hampel1974influence} can be introduced. If we define $\hat{\theta}_{+}$ as an an estimator of the same form of $\hat{\theta}$ based on the same $n$ observations, but also on an additional observation $z_{o}$, a first finite sample version of Hampel's influence function  can be defined as $I_{+}\left(\theta, z_{o}\right)=(n+1)\left(\hat{\theta}_{+}-\hat{\theta}\right)$. As an alternative we could use the expression  $I_{-}\left(\theta, z_{o}\right)=(n-1)\left(\hat{\theta}_{-}-\hat{\theta}\right)$ where $\hat{\theta}_{-}$ is an estimator of the same form of $\hat{\theta}$ obtained by eliminating one of the original observations.
Given the previous definitions, it is clear that, both of these quantities, in general, depend only on the amount of contamination $z_{o}$. 

However, when considering data distributed in space, the definition of the influence function is different than in the generality of cases because the number of locations is given (e. g. the number of pixels of an image or the number of regions within one country) so that the influence cannot be measured by adding nor subtracting one of the existing units, rather by contaminating one of them.
Following this approach, we will indicate with:

\begin{equation}
I_{cont}\left(\theta\right)=n\left(\hat{\theta}_{cont}-\hat{\theta}\right)
\end{equation}

the influence function of the parameter $\theta$ after the contamination one of the existing units, with $\hat{\theta}_{cont}$ representing the estimator of the parameter after such a contamination. 
Intuitively, due to the nature of dependence between spatial data, $I_{cont}\left(\theta\right)$ now does not depend only on the contaminated value, but also on (i) the location of the contaminated unit, (ii) on its connection with the neighbouring locations, and (iii) on the values observed in the neighbouring locations. 
Indeed, if if the contaminated location $i$ is strongly connected with other locations (that is, it is a $dominant$ $unit$ according to the definition of \cite{pesaran2020econometric}), the influence of the contaminated value will be stronger than in the case of loosely connected units because in this case its effect propagates also to the neighboring units and hence it corrupts more substantially the spatial correlation parameters.

In the case of MC and APLE coefficients, some theoretical results could be derived to support such an intuition. At the basis of both MC and APLE statistics, indeed, is the calculation of the spatial autocovariance appearing in their numerator, say $\gamma=n^{-1} Z^{T} W Z$ (see Equations (2) and (4)).

Let us consider again a set of $n$ observations $\left(z_{1}, z_{2}, \ldots, z_{n}\right.$, ) centered around their mean and let us further assume, without losing generality, that we contaminate the first unit and that this unit before the contamination had a value equal to zero.

Consequently, the mean after contamination, will be $\bar{z}_{\text {cont }}=\frac{z_{1}}{n}$.

Let us now consider the empirical spatial autocovariance after the perturbation of unit 1. With the symbolism already introduced, this term can be written as:

\begin{equation}
\hat{\gamma}_{\text {cont }}=n^{-1}Z^{T} L(Z)=n^{-1} \sum_{i=1}^{n} \sum_{j=1}^{n} w_{i j} z_{i} z_{j}-\left(\frac{z_{1}}{n}\right)^{2}
\end{equation}

which can be also expressed as:

\begin{equation}
\begin{aligned}
& \hat{\gamma}_{\text {cont }}=n^{-1} \sum_{i=1}^{n} z_{i} \sum_{j=1}^{n} w_{i j} z_{j}-\left(\frac{z_{1}}{n}\right)^{2}=n^{-1}\left(z_{1}+\sum_{i=2}^{n} z_{i}\right)\left[\sum_{j=1}^{n} w_{i j} z_{j}\right]-\left(\frac{z_{1}}{n}\right)^{2}= \\
& =n^{-1}\left(\sum_{i=2}^{n} \sum_{j=2}^{n} w_{i j} z_{i} z_{j}+z_{1} \sum_{i=2}^{n} w_{i 1} z_{i}+z_{1} \sum_{j=2}^{n} w_{1 j} z_{j}\right)-\left(\frac{z_{1}}{n}\right)^{2}
\end{aligned}
\end{equation}

$w_{i 1}$ denoting the first row of the matrix $\mathrm{W}$ and $w_{1 j}$ its first column. Having assumed that before the perturbation $z_{1}=0$, we have, that, before the perturbation, the spatial autocovariance can be written as:

\begin{equation}
\hat{\gamma}=n^{-1} \sum_{i=2}^{n} \sum_{j=2}^{n} w_{i j} z_{i} z_{j}
\end{equation}

From the definition given in Equation (6), the empirical influence function associated to the spatial autocovariance can be written as:

\begin{equation}
\hat{\gamma}_{\text {cont }}=n^{-1} Z^T L(Z)=n^{-1} \sum_{i=1}^n \sum_{j=1}^n w_{i j} z_i z_j-\left(\frac{z_1}{n}\right)^2
\end{equation}

In the previous equation, the term $\sum_{i=2}^{n} w_{i 1} z_{i}$ represents the spatial lag of $z_{1}$ (that is the average of the values observed in the neighbourhood of the perturbated unit 1), while the term $\sum_{j=2}^{n} w_{1 j} z_{j}$ represents the contribution of those sites that admit unit 1 as one of their neighbours. To give a flavour of the factors that may determine the influence function, let us consider a specific case when Equation (10) can be simplified. Indeed, if we assume that the $n$ locations are laid on a regular square lattice grid mapped onto a torus with $\eta_{i}=\eta \forall i$, the $\mathrm{W}$ matrix maintains its original property of symmetry even after the row standardization. Hence $\sum_{i=2}^{n} w_{i 1} z_{i}=\sum_{j=2}^{n} w_{1 j} z_{j}$ so that Equation (9) simplifies into:

\begin{equation}
    I_{cont}\left(\gamma\right)=2 z_{1} \sum_{i=2}^{n} w_{i 1} z_{i}-\left(\frac{z_{1}}{n}\right)^{2}
\end{equation}

Equation (11) represents a parabola with lower convexity passing through the origin, thus showing that:

\begin{enumerate}
    \item When $z_{1}=0, I_{\text {cont }}=0$ as it should be expected
    \item $\quad I_{\text {cont }}$ reaches its maximum at $z_{1}=n \sqrt{2 \sum_{i=2}^{n} w_{i 1} z_{i}}$
with a rate that is emphasized by the term $2 \sum_{i=2}^{n} w_{i 1} z_{i}$
    \item When $n \rightarrow \infty$ the empirical influence function degenerates to a straight line which increases monotonically passing through the origin, with a slope given by $2 \sum_{i=2}^{n} w_{i 1} z_{i}$. Such a term depends on the number of neighbours of the perturbated unit. Hence, in different weighting schemes, $I_{c o n t}$, will be related to $\eta_{i}$
\end{enumerate}

To evaluate the relative robustness to outliers of the various spatial correlation measures, we run a simulation exercise by injecting different levels of contamination (with  $-10 \leq z_1 \leq 10$) in a randomly chosen spatial unit of a 10-by-10 regoular square lattice grid. The starting dataset is generated from a standardized Gaussian spatial autoregressive model (\cite{cressie}) characterized by a spatial correlation parameter equal to 0.5 and using the rook's case neighbourhood. The results are reported in Figure 1 which illustrates the behaviour of the simulated influence function evaluated by averaging the results of 1,000 simulation runs. Notably, the empirical influence functions of MC and GC perfectly overlap, coherently with the fact that, as said, they are inferentially equivalent. In contrast, APLE is considerably less robust to the presence of outliers with respect to MC and GC.

\begin{figure}[h!]
    \centering
    \includegraphics[width=\textwidth]{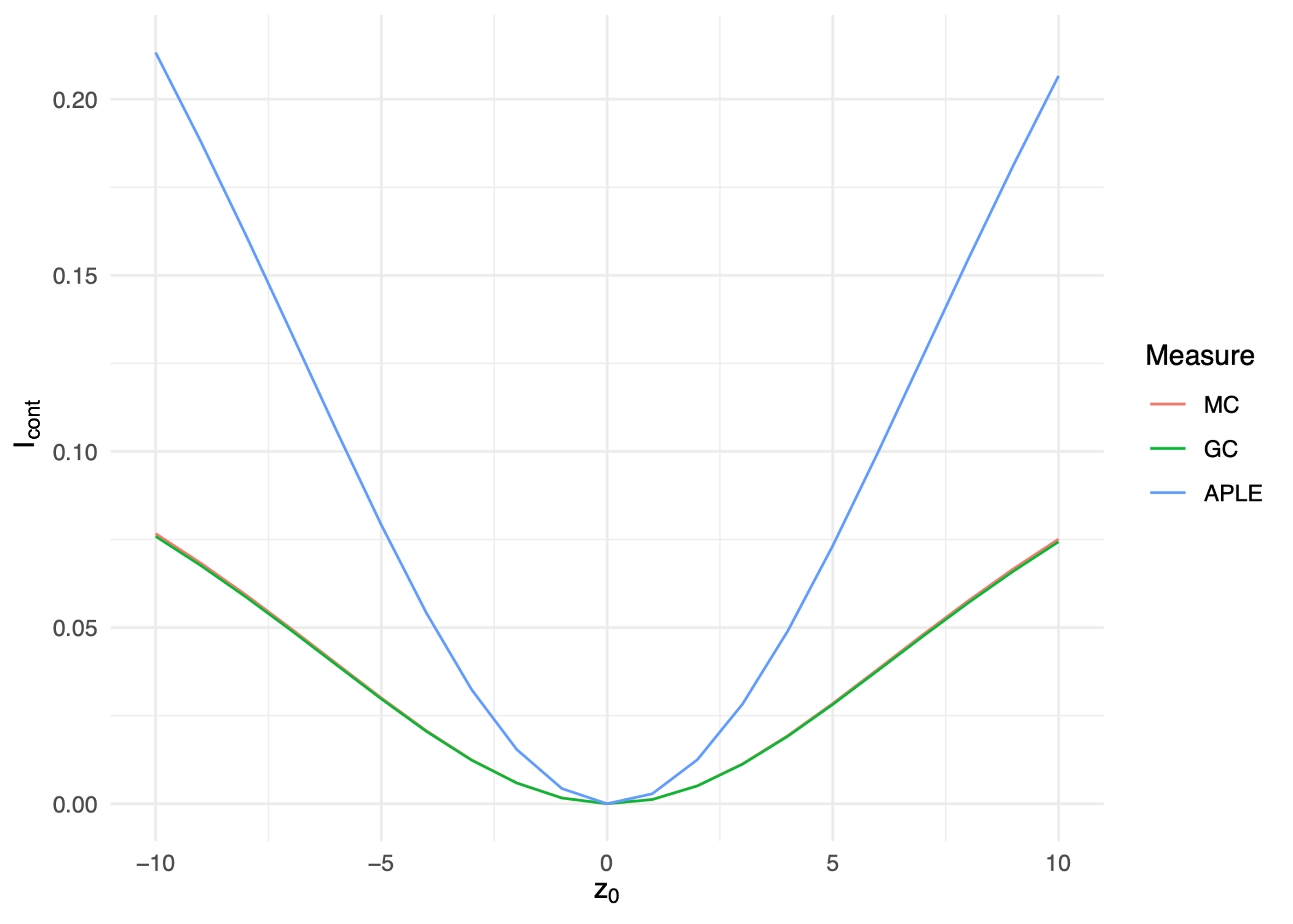}
    \caption{Simulated $I_{cont}$ influence function for Moran Coefficient (MC), Geary's Coefficient (GC), and APLE }
    \label{fig:influence_function} 
\end{figure}

The findings reported in this section provide scope for methods to robustizing the traditional
spatial correlation measures. This task will be  pursued in the next section.

\section{Robust estimation of spatial correlation}
\
\subsection{Specification of robust estimators}
The unboundedness of the influence function of the three traditional spatial correlation measures discussed in the previous section, leads us to the search of new, empirically designed,  \textit{ad hoc} robust estimators to protect against the potentially severe consequences due to the presence of abnormal values.

Preliminarily to the presentation of the various estimators, let us first introduce the notion of the robust spatial lag (RL), defined as the weighted median of the neighbours of $x_{i}$ according to the topology described by the matrix $W$.

\begin{equation}
RL\left(x_{i}\right)=\operatorname{Med}\left(x_{j}\right) ; j \in N(i)
\label{eq:RL}
\end{equation}

Notice that, although substantially different, the notion of the robust spatial lag shares some resemblance to the idea of the spatial median (see \cite{gentleman1966robust}, \cite{brown1983statistical} and \cite{brown1997effect}). 

Given the above definition, we can then consider the following alternative robust spatial correlation measures to be used in hypothesis testing:

a) MC using the robust spatial lag definition in place of the spatial lag definition:

\begin{equation}
MC=\frac{Z^{T} RL(Z)}{Z^{T} Z}
\label{eq:RMC}
\end{equation}

b) APLE using the robust spatial lag definition in place of the spatial lag definition:

\begin{equation}
R A P L E=\frac{1}{2} \frac{\left[R L(Z)^{T} Z+Z^{T} R L(Z)\right]}{R L(Z)^{T} R L(Z)+\operatorname{tr}\left(W^{2}\right) Z^{T} Z / n}
\label{eq:RAPLE}
\end{equation}

c) GC using robust versions of the two sums of squares appearing in the numerator and respectively in the denominator of Equation (4):

\begin{equation}
R G C=\frac{(2 n \bar{\eta})^{-1} \sum_{i=1}^{n} \sum_{j=1}^{n} w_{i j}\left|z_{i}-z_{j}\right|}{(n-1)^{-1} \sum_{i=1}^{n}\left|x_{i}\right|}
\label{eq:RGC}
\end{equation}

To introduce a further measure, let us recall the general notion of robust correlation proposed by \cite{gnanadesikan1972robust}:

\begin{equation}
\frac{S(a X+b Y)^{2}-S(a X-b Y)^{2}}{S(a X+b Y)^{2}+S(a X-b Y)^{2}}
\label{eq:GKintro}
\end{equation}
with $a=S(X)^{-1}, b=S(Y)^{-1}$, and $\mathrm{S}$ any robust measure of scale.

d)  If $X$ is substituted by $Z, Y$ by the spatially lagged value of $Z$ and we opt the Median Absolute Deviation from the median (MAD) as a robust measure of scale, we obtain a further alternative as:

\begin{equation}
G K=\frac{M A D(a Z+L(Z))^{2}-M A D(a Z-b L(Z))^{2}}{M A D(a Z+b L(Z))^{2}+M A D(a Z-b L(Z))^{2}}
\label{eq:GK}
\end{equation}

e) Finally, the last measure can be further robustized by using the robust spatial lag definition:

\begin{equation}
G K 2=\frac{M A D(a Z+L R(Z))^{2}-M A D(a Z-b R L(Z))^{2}}{M A D(a Z+b R L(Z))^{2}+M A D(a Z-b R L(Z))^{2}}
\label{eq:GK2}
\end{equation}

\subsection{Monte Carlo study of the robust estimators}
The finite sample properties of all the suggested robust measures of spatial correlation will be now investigated through a series of Monte Carlo experiments. First, adopting the same strategy described in Section 3 for the traditional spatial correlation measures, we will derive the simulated empirical influence functions of the various measures obtained by contaminating one of the existing spatial units (Section 4.2.1). Secondly, we will investigate, in different experimental situations, the empirical power of the tests of hypothesis associated with the robust measures under the null of no spatial correlation (Section 4.2.2).

In Figure 2, we report the simulated influence functions for the robust spatial correlation alternatives reported in Equations (13) through (18). Lacking the exact distribution of the various measures, and to reduce any potential bias in the comparison caused by approximation errors, the power of the various tests is evaluated numerically by analysing the results of 1,000 Monte Carlo runs in each experiment.
The consistent use of such an approach guarantees full comparability across the various analyses.

 When contrasted with the original measures presented in Figure 1, all the measures demonstrate enhanced robustness. Specifically, while RAPLE is notably the most sensitive to outliers, both RMC and RGC exhibit comparable resilience to anomalous values. Amongst all, our proposed measures, GK and GK2, stand out; however, GK2 might be overly robust, potentially indicating a diminished sensitivity.

\begin{figure}[h!]
    \centering
    \includegraphics[width=\textwidth]{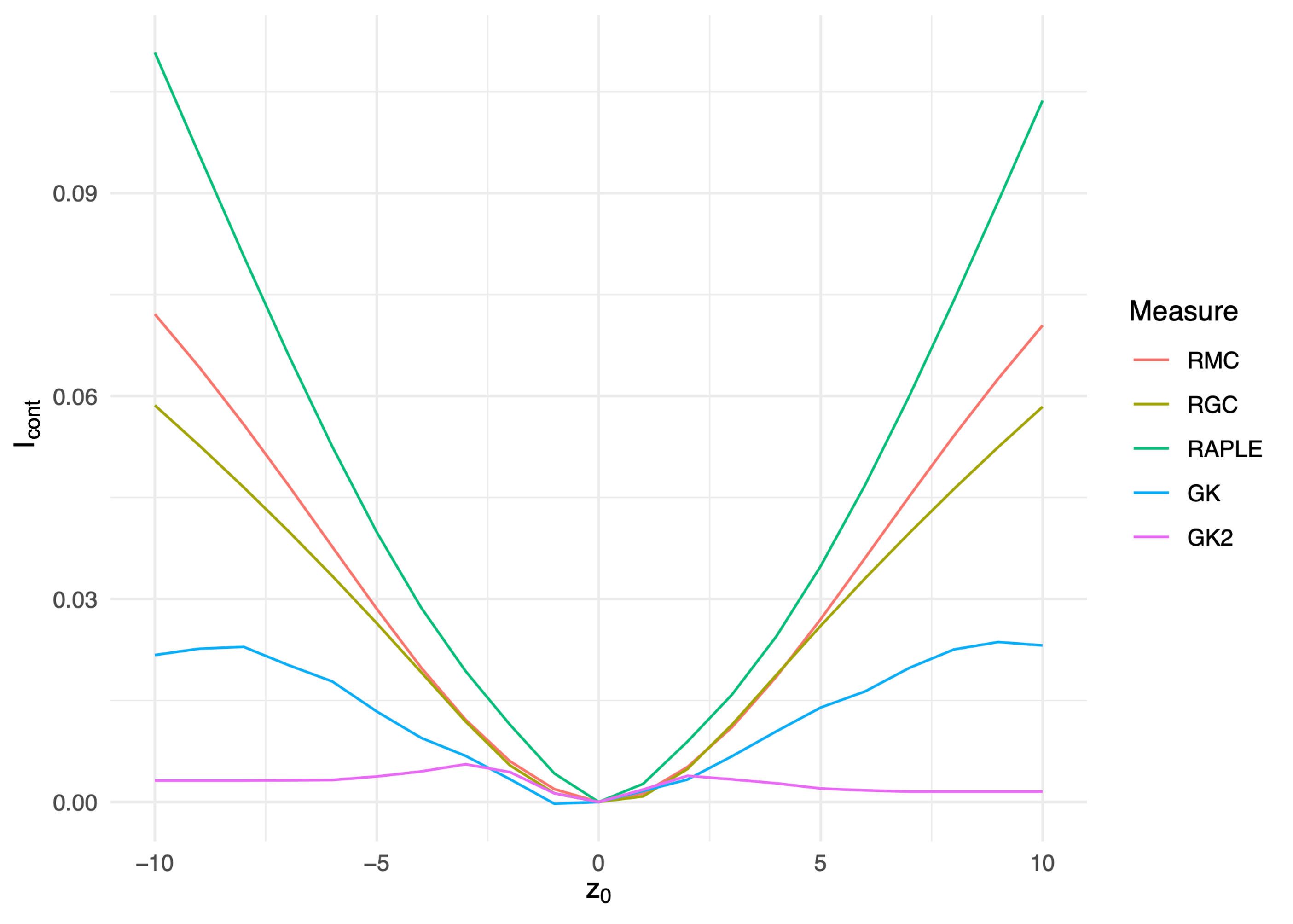}
    \caption{Simulated influence function for Robust Moran Coefficient (RMC), Robust Geary's Coefficient (RGC), Robust APLE (RAPLE), GK and GK2 Coefficients.}
    \label{fig:influence_function2} 
\end{figure}

In the light of the previous results, the performance of the various proposed estimators will be now studied through a simulation based on data laid on two regoular square lattice grids of dimension $n=100$ (10-by-10) and a $n=400 $ (20-by-20) respectively. The data are originally generated with the following spatial autoregressive model \citep{cressie1993aggregation}:

\begin{equation}
Z=\rho W Z+\epsilon
\end{equation}

We considered seven values of the parameter $\rho$ ($\rho= -0.7, -0.5, -0.3, 0, 0.3, 0.5, 0.7)$ and, following \cite{devlin1975robust}, four different distributions for the disturbance term $\epsilon$, namely: 

\begin{enumerate}[label=\alph*)]
    \item Standardized Normal distribution 
    \item Cauchy distribution with location parameter 0 and scale parameter 1
    \item Laplace distribution with location parameter 0 and scale parameter 1
    \item A mixture of two Normal distributions with unitary variance and different expected values $[0.95 N(0, 1)+0.05 N(3,1)]$
\end{enumerate}

It is essential to highlight that, although strongly kurtotic, both Cauchy and Laplace distributions are symmetric and centered, while the fourth one (the mixture of two normal distributions) is intrinsically asymmetric. The inclusion of this further typology represents a novel approach as it diverges from traditional testing methods in this area \citep{gnanadesikan1972robust}. Indeed, the presence of marked skewness and kurtosis is particularly pronounced in many real-world datasets so that, by incorporating in our test both of them, we present a more comprehensive and realistic exam which recognizes the complexities present in actual data.

Furthermore, to explore the effects of different connectivity in the spatial system, we also considered two definitions of the $W$ matrix using the rook's case definition (where $\eta_{i}=4$ for each $i$ apart from those located at the edge of the lattice grid) and the queen's case (where $\eta_{i}=8$ for each $i$ apart from those located at the edge of the lattice grid) \citep{cressie1993aggregation}. We repeated each experimental situation 1,000 times.

To study the relative performance of each of the proposed estimators, Table 1 displays the empirical power of the test obtained through our Monte Carlo simulations in each of the experimental situations.

\begin{table}[ht]
\centering
\caption{Empirical power of the various spatial correlation tests at $\rho=0$. (Percentage of simulation cases when the null of no spatial autocorrelation is rejected at the nominal size of 0.05). In the simulations we set $n=100$ and we employed the Queen neighbourhood for the W matrix.\\}
\begin{tabular}{rrlrrrr}
  \hline
 Measure & Normal & Cauchy & Laplace & Mixture \\ 
  \hline
  MC & 0.06 & 0.07 & 0.05 & 0.54 \\ 
   GC & 0.05 & 0.06 & 0.05 & 0.77 \\
 APLE & 0.06 & 0.06 & 0.05 & 0.55 \\ 
  RMC & 0.06 & 0.06 & 0.05 & 0.34 \\ 
   RGC & 0.05 & 0.06 & 0.05 & 0.58 \\ 
    RAPLE & 0.06 & 0.05 & 0.04 & 0.35 \\ 

 GK & 0.05 & 0.05 & 0.05 & 0.09 \\ 
 GK2 & 0.05 & 0.05 & 0.05 & 0.07 \\ 
 
   \hline
\end{tabular}
\end{table}

The Online Appendix provides a comprehensive list of all simulation results. Observing the results obtained for the Normal, Cauchy, and Laplace distributions in Table 1, it is evident how, at $\rho=0$, all measures show an empirical power hovering around the the nominal value of 0.05. However, when analyzing the Mixture distribution, a distinct pattern emerges. Indeed, in this case, every measure departs markedly from the target 0.05 value. This indicates a significant susceptibility to the asymmetry and an inherent complexities of data generated with the mixture distribution. Notably, while this happens for most measures, the proposed measures GK and GK2, display remarkable resilience. Their superior performance over the other measures underscores the potential and efficacy of our proposed approach. This shows not only the robustness of our proposed measures, but also their superiority in terms of adaptability to diverse data distributions, particularly those that are skewed or mixed.

In Figure 3, we present the empirical power at $\rho=0$ calculated in two different weight matrices scenarios: the one based on the Rook's case neighbourhood and the one based on the Queen's case. The visual inspection of the graph indicates that for the Normal, the Cauchy and the Laplace distributions, all measures consistently exhibit robustness to outliers (with an empirical power close to the nominal value of 0.05) irrespective of the weight matrices chosen. However, in the case of the Mixture distribution, GK and GK2 stand out as significantly more robust than the other alternatives.

\begin{figure}[h!]
    \centering
    \includegraphics[width=\textwidth]{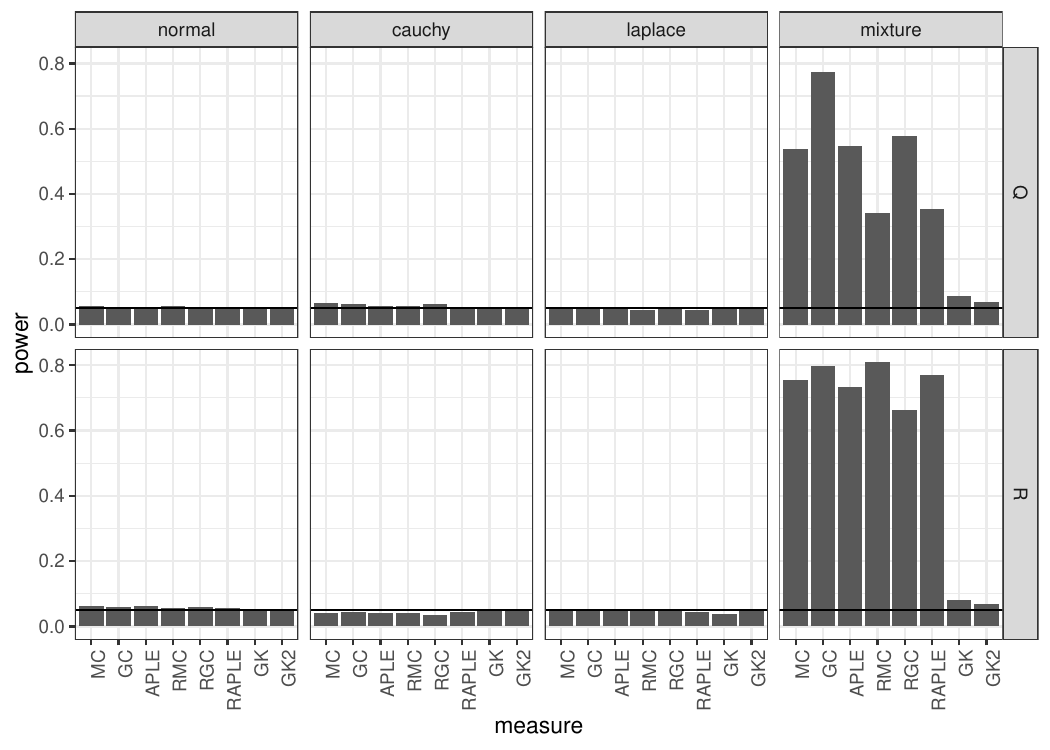}
    \caption{Simulated empirical power of the various measures for $\rho=0$, $n=100$ for queen's (Q) and rook's (R) cases Weight Matrices.}
    \label{fig:power1} 
\end{figure}

The enhanced power of the mixture in the rook's case, when compared to the queen's case, is to be attributed to the variations in the associated connectivity. Indeed, when the number of neighbors increases, as it happens in the queen's case scenario, the power of the test diminishes. This observation aligns with the previous findings reported in Section 3.

In Figure 4, we show the empirical power at $\rho=0$ calculated for each distribution, comparing the results obtained with different sample dimensions when n = 100 and when n = 400. Apart from the intuitive higher power in larger samples, the patterns displayed in Figure 4 are strikingly similar to those presented in Figure 3, reinforcing the consistency and reliability of our findings across different metrics and evaluations. 

More details on the simulation results, are contained in the online Appendix which, in particular, reports the empirical power function (evaluated at $\rho= -0.7, -0.5, -0.3, 0, 0.3, 0.5, 0.7)$ for all measures and all simulation scenarios. Notice that we considered two sample sizes and two weighting matrices as described above, but we limited our analysis to only the queen's case W matrix when n = 400.

In general, we observe that the Geary coefficient performances are always the poorest with respect to all other measures and, furthermore, that the GK measure always outperforms GK2 in terms of the power.
Analysing the results into detail, we observe that, when the disturbances in the autoregression model (19) are normally distributed, both MC and APLE outperform their robust alternatives (and in particular GK and GK2) irrespective of the sample size and of the weight matrix chosen. So, in this case, no need of correction emerges.

Different results are obtained when considering non-normally distributed residuals.
Indeed, in the case of residuals that are non-normal, but still symmetrically distributed (as it happens in the case of the Cauchy and of the Laplace distributions) results still share a certain resemblance with those observed in the Normal case in accordance with the findings of \cite{griffith2010moran} with some remarkable difference between them.
In fact, in the case of the Laplace distribution, both MC and APLE are still the best choice: they dominate GK and GK2 and they don't get much benefit from the robust lag definition correction. In contrast, in the case of residuals distributed following the Cauchy law, the performances of GK are comparable with those obtained using MC, while the best performing measures appear to be RMC and RAPLE.
Finally, when regression residuals in Equation (19) are generated by a mixture of two normal distributions (thus introducing a greater degree of skewness, GK (and to a lesser extent GK2) are uniformly most powerful than the other alternatives.

\begin{figure}[h!]
    \centering
    \includegraphics[width=\textwidth]{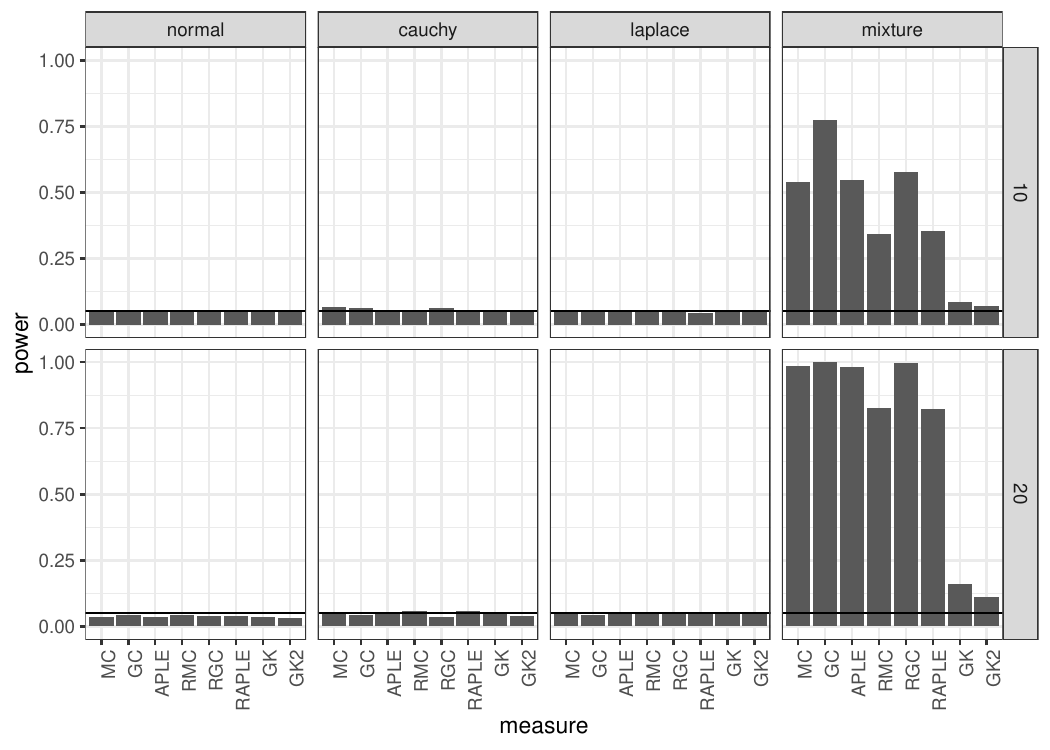}
    \caption{Simulated empirical power of the various measures for $\rho=0$, Queen (Q) for $n=100$ and $n=400$.}
    \label{fig:power2} 
\end{figure}

\section{Concluding remarks}

In this research, we ventured into the domain of spatial correlation measures, aiming at introducing methods to robustize the traditional metrics. Our approach is grounded in the traditional examination of influence functions, as delineated by \cite{hampel1974influence}. This analysis granted us the opportunity to evaluate the relative robustness of conventional metrics with respect to our alternative proposals.

Leveraging on the capabilities of Monte Carlo experiments, we simulated diverse spatial data sets where we incorporated different outlier scenarios. Such simulations were pivotal in discerning the efficacy of our proposed measures. 

In the case of normally distributed data, no correction is necessary because all traditional measure are still robust to the presence of outliers. In the case of Cauchy and Laplace distributions (which are still symmetric but with a strong kurtosis), in accordance with the findings of \cite{griffith2010moran}, MC and APLE are still relatively robust. In particular, while in the Laplace distribution case we do not notice any remarkable differences from the Normal case, in the case of the Cauchy distribution both MC and APLE improve their performances if we replace the traditional spatial lag definition with the notion of robust lag. Finally, in stark contrast, when data are generated with a strong skew  (like in a Mixture distribution), we observe a more convoluted scenario where the GK metrics (and to a lesser extent GK2) stand out as exemplary in their resilience to the presence of abnormal values.

In conclusion, the implications of our findings are profound. The GK measure, with its inherent robustness, usher in a fresh perspective to spatial data analysis when they depart markedly from the Gaussian case in terms of the skew. Indeed, in many empirical circumstances spatial data are characterized by this form of non-normality as it happens, for instance, when dealing with socio-economic spatial data (e. g. per capita income, population density and many other), environmental and epidemiological data to name only a few. It should be emphasized that the empirical situations indicated above are precisely those cases when a spatial correlation analysis is often more necessary. These findings emphasize the continuous need for innovative methods that are both robust and adaptive to the ever-evolving challenges of spatial data.

Furthermore, this foundational study paves the way to evaluate the influence function across spatial dimensions. Such an approach promises insights into the impact of observed values, taking into consideration not just the magnitude, but also the spatial location of influential observations. This spatially-aware influence function evaluation has the potential to significantly enriching our knowledge in the area of spatial statistics and econometrics and to the understanding of spatial data patterns and dynamics.

\newpage
\bibliography{bibliography.bib}

\newpage
\section*{Appendix}

Table A.1: Empirical power function for Normal distribution (n=100) and Queen (Q) or Rook (R) W matrices
\begin{table}[ht]
\centering
\begin{tabular}{lrlrrrrrrr}
  \hline
\multirow{2}{*}{Measure} & \multirow{2}{*}{n} & \multirow{2}{*}{$W$} & \multicolumn{7}{c}{$\rho$} \\
\cline{4-10}
 & & & -0.7 & -0.5 & -0.3 & 0 & 0.3 & 0.5 & 0.7 \\
\hline
MC &  10 & Q & 0.96 & 0.78 & 0.43 & 0.06 & 0.54 & 0.90 & 1.00 \\ 
  MC &  10 & R & 1.00 & 0.99 & 0.73 & 0.06 & 0.73 & 0.99 & 1.00 \\ 
  GC &  10 & Q & 0.55 & 0.39 & 0.22 & 0.05 & 0.43 & 0.83 & 0.99 \\ 
  GC &  10 & R & 1.00 & 0.99 & 0.73 & 0.06 & 0.73 & 0.98 & 1.00 \\ 
  APLE &  10 & Q & 0.96 & 0.78 & 0.42 & 0.06 & 0.54 & 0.90 & 1.00 \\ 
  APLE &  10 & R & 1.00 & 0.99 & 0.73 & 0.06 & 0.74 & 0.99 & 1.00 \\ 
  RMC &  10 & Q & 0.88 & 0.66 & 0.34 & 0.06 & 0.49 & 0.88 & 0.99 \\ 
  RMC &  10 & R & 1.00 & 0.98 & 0.68 & 0.06 & 0.69 & 0.98 & 1.00 \\ 
  RGC &  10 & Q & 0.85 & 0.62 & 0.33 & 0.05 & 0.48 & 0.87 & 0.99 \\ 
  RGC &  10 & R & 1.00 & 0.96 & 0.65 & 0.06 & 0.66 & 0.97 & 1.00 \\ 
  RAPLE &  10 & Q & 0.89 & 0.68 & 0.35 & 0.06 & 0.49 & 0.88 & 0.99 \\ 
  RAPLE &  10 & R & 1.00 & 0.99 & 0.69 & 0.06 & 0.69 & 0.98 & 1.00 \\ 
  GK &  10 & Q & 0.78 & 0.51 & 0.25 & 0.05 & 0.31 & 0.68 & 0.94 \\ 
  GK &  10 & R & 0.99 & 0.86 & 0.50 & 0.05 & 0.48 & 0.89 & 0.99 \\ 
  GK2 &  10 & Q & 0.60 & 0.38 & 0.21 & 0.05 & 0.28 & 0.58 & 0.90 \\ 
  GK2 &  10 & R & 0.99 & 0.78 & 0.44 & 0.05 & 0.41 & 0.81 & 0.98 \\ 
   \hline
\end{tabular}
\end{table}

\newpage
Table A.2: Empirical power function for Cauchy distribution (n=100) and Queen (Q) or Rook (R) W matrices
\begin{table}[ht]
\centering
\begin{tabular}{lrlrrrrrrr}
  \hline
\multirow{2}{*}{Measure} & \multirow{2}{*}{n} & \multirow{2}{*}{$W$} & \multicolumn{7}{c}{$\rho$} \\
\cline{4-10}
 & & & -0.7 & -0.5 & -0.3 & 0 & 0.3 & 0.5 & 0.7 \\
\hline
MC &  10 & Q & 0.99 & 0.91 & 0.64 & 0.06 & 0.78 & 0.97 & 1.00 \\ 
  MC &  10 & R & 1.00 & 0.99 & 0.89 & 0.04 & 0.89 & 0.99 & 1.00 \\ 
  GC &  10 & Q & 0.76 & 0.61 & 0.32 & 0.06 & 0.47 & 0.93 & 0.99 \\ 
  GC &  10 & R & 1.00 & 0.99 & 0.84 & 0.04 & 0.83 & 0.99 & 1.00 \\ 
  APLE &  10 & Q & 0.99 & 0.91 & 0.62 & 0.06 & 0.78 & 0.98 & 1.00 \\ 
  APLE &  10 & R & 1.00 & 0.99 & 0.88 & 0.04 & 0.89 & 1.00 & 1.00 \\ 
  RMC &  10 & Q & 1.00 & 1.00 & 0.96 & 0.06 & 0.96 & 1.00 & 1.00 \\ 
  RMC &  10 & R & 1.00 & 1.00 & 0.97 & 0.04 & 0.97 & 1.00 & 1.00 \\ 
  RGC &  10 & Q & 0.19 & 0.16 & 0.11 & 0.06 & 0.75 & 1.00 & 1.00 \\ 
  RGC &  10 & R & 0.98 & 0.91 & 0.70 & 0.04 & 0.93 & 1.00 & 1.00 \\ 
  RAPLE &  10 & Q & 1.00 & 0.99 & 0.96 & 0.05 & 0.97 & 1.00 & 1.00 \\ 
  RAPLE &  10 & R & 1.00 & 1.00 & 0.98 & 0.04 & 0.98 & 1.00 & 1.00 \\ 
  GK &  10 & Q & 0.99 & 0.92 & 0.63 & 0.05 & 0.73 & 0.97 & 1.00 \\ 
  GK &  10 & R & 1.00 & 0.99 & 0.81 & 0.05 & 0.79 & 0.98 & 1.00 \\ 
  GK2 &  10 & Q & 0.63 & 0.42 & 0.26 & 0.05 & 0.74 & 0.96 & 1.00 \\ 
  GK2 &  10 & R & 1.00 & 0.98 & 0.74 & 0.05 & 0.69 & 0.97 & 1.00 \\ 
   \hline
\end{tabular}
\end{table}

\newpage
Table A.3: Empirical power function for Laplace distribution (n=100) and Queen (Q) or Rook (R) W matrices
\begin{table}[ht]
\centering
\begin{tabular}{lrlrrrrrrr}
  \hline
\multirow{2}{*}{Measure} & \multirow{2}{*}{n} & \multirow{2}{*}{$W$} & \multicolumn{7}{c}{$\rho$} \\
\cline{4-10}
 & & & -0.7 & -0.5 & -0.3 & 0 & 0.3 & 0.5 & 0.7 \\
\hline
MC &  10 & Q & 0.97 & 0.78 & 0.43 & 0.05 & 0.57 & 0.90 & 1.00 \\ 
  MC &  10 & R & 1.00 & 0.99 & 0.73 & 0.05 & 0.72 & 0.98 & 1.00 \\ 
  GC &  10 & Q & 0.57 & 0.38 & 0.20 & 0.05 & 0.42 & 0.83 & 0.99 \\ 
  GC &  10 & R & 1.00 & 0.99 & 0.73 & 0.05 & 0.72 & 0.98 & 1.00 \\ 
  APLE &  10 & Q & 0.96 & 0.78 & 0.44 & 0.05 & 0.57 & 0.91 & 1.00 \\ 
  APLE &  10 & R & 1.00 & 0.99 & 0.74 & 0.05 & 0.72 & 0.98 & 1.00 \\ 
  RMC &  10 & Q & 0.94 & 0.74 & 0.42 & 0.05 & 0.60 & 0.92 & 1.00 \\ 
  RMC &  10 & R & 1.00 & 0.99 & 0.73 & 0.05 & 0.75 & 0.99 & 1.00 \\ 
  RGC &  10 & Q & 0.85 & 0.63 & 0.34 & 0.05 & 0.54 & 0.91 & 1.00 \\ 
  RGC &  10 & R & 1.00 & 0.97 & 0.69 & 0.05 & 0.70 & 0.98 & 1.00 \\ 
  RAPLE &  10 & Q & 0.96 & 0.78 & 0.45 & 0.04 & 0.61 & 0.93 & 1.00 \\ 
  RAPLE &  10 & R & 1.00 & 0.99 & 0.75 & 0.04 & 0.75 & 0.99 & 1.00 \\ 
  GK &  10 & Q & 0.87 & 0.62 & 0.31 & 0.05 & 0.42 & 0.81 & 0.98 \\ 
  GK &  10 & R & 1.00 & 0.92 & 0.56 & 0.04 & 0.56 & 0.94 & 1.00 \\ 
  GK2 &  10 & Q & 0.66 & 0.43 & 0.24 & 0.05 & 0.37 & 0.73 & 0.96 \\ 
  GK2 &  10 & R & 1.00 & 0.87 & 0.50 & 0.05 & 0.51 & 0.87 & 0.99 \\ 
   \hline
\end{tabular}
\end{table}

\newpage
Table A.4: Empirical power function for Mixture distribution (n=100) and Queen (Q) or Rook (R) W matrices
\begin{table}[ht]
\centering
\begin{tabular}{lrlrrrrrrr}
  \hline
\multirow{2}{*}{Measure} & \multirow{2}{*}{n} & \multirow{2}{*}{$W$} & \multicolumn{7}{c}{$\rho$} \\
\cline{4-10}
 & & & -0.7 & -0.5 & -0.3 & 0 & 0.3 & 0.5 & 0.7 \\
\hline
MC &  10 & Q & 0.61 & 0.19 & 0.04 & 0.54 & 0.95 & 1.00 & 1.00 \\ 
  MC &  10 & R & 1.00 & 0.70 & 0.10 & 0.75 & 1.00 & 1.00 & 1.00 \\ 
  GC &  10 & Q & 0.02 & 0.01 & 0.00 & 0.77 & 0.98 & 1.00 & 1.00 \\ 
  GC &  10 & R & 0.99 & 0.66 & 0.09 & 0.80 & 1.00 & 1.00 & 1.00 \\ 
  APLE &  10 & Q & 0.60 & 0.19 & 0.04 & 0.55 & 0.96 & 1.00 & 1.00 \\ 
  APLE &  10 & R & 0.99 & 0.67 & 0.08 & 0.74 & 1.00 & 1.00 & 1.00 \\ 
  RMC &  10 & Q & 0.68 & 0.38 & 0.14 & 0.34 & 0.86 & 0.99 & 1.00 \\ 
  RMC &  10 & R & 0.99 & 0.64 & 0.10 & 0.81 & 1.00 & 1.00 & 1.00 \\ 
  RGC &  10 & Q & 0.29 & 0.07 & 0.02 & 0.58 & 0.96 & 1.00 & 1.00 \\ 
  RGC &  10 & R & 0.98 & 0.63 & 0.09 & 0.66 & 1.00 & 1.00 & 1.00 \\ 
  RAPLE &  10 & Q & 0.68 & 0.39 & 0.15 & 0.36 & 0.87 & 0.99 & 1.00 \\ 
  RAPLE &  10 & R & 0.98 & 0.59 & 0.07 & 0.77 & 1.00 & 1.00 & 1.00 \\ 
  GK &  10 & Q & 0.72 & 0.41 & 0.20 & 0.09 & 0.47 & 0.80 & 0.96 \\ 
  GK &  10 & R & 0.99 & 0.83 & 0.38 & 0.08 & 0.59 & 0.93 & 0.99 \\ 
  GK2 &  10 & Q & 0.56 & 0.32 & 0.16 & 0.07 & 0.35 & 0.69 & 0.93 \\ 
  GK2 &  10 & R & 0.97 & 0.74 & 0.31 & 0.07 & 0.49 & 0.85 & 0.99 \\ 
   \hline
\end{tabular}
\end{table}

\newpage

Table A.5: Empirical power function for Normal distribution (n=400) and Queen (Q) W matrix
\begin{table}[ht]
\centering
\begin{tabular}{lrlrrrrrrr}
  \hline
\multirow{2}{*}{Measure} & \multirow{2}{*}{n} & \multirow{2}{*}{$W$} & \multicolumn{7}{c}{$\rho$} \\
\cline{4-10}
 & & & -0.7 & -0.5 & -0.3 & 0 & 0.3 & 0.5 & 0.7 \\
\hline
MC &  20 & Q & 1.00 & 1.00 & 0.91 & 0.03 & 0.95 & 1.00 & 1.00 \\ 
  GC &  20 & Q & 0.99 & 0.91 & 0.56 & 0.04 & 0.85 & 1.00 & 1.00 \\ 
  APLE &  20 & Q & 1.00 & 1.00 & 0.91 & 0.04 & 0.96 & 1.00 & 1.00 \\ 
  RMC &  20 & Q & 1.00 & 0.99 & 0.82 & 0.04 & 0.93 & 1.00 & 1.00 \\ 
  RGC &  20 & Q & 1.00 & 0.99 & 0.80 & 0.04 & 0.93 & 1.00 & 1.00 \\ 
  RAPLE &  20 & Q & 1.00 & 0.99 & 0.82 & 0.04 & 0.93 & 1.00 & 1.00 \\ 
  GK &  20 & Q & 1.00 & 0.97 & 0.67 & 0.04 & 0.77 & 1.00 & 1.00 \\ 
  GK2 &  20 & Q & 0.98 & 0.86 & 0.49 & 0.03 & 0.69 & 0.98 & 1.00 \\ 
   \hline
\end{tabular}
\end{table}

\newpage
Table A.6: Empirical power function for Cauchy distribution (n=400) and Queen (Q) W matrix
\begin{table}[ht]
\centering
\begin{tabular}{lrlrrrrrrr}
  \hline
\multirow{2}{*}{Measure} & \multirow{2}{*}{n} & \multirow{2}{*}{$W$} & \multicolumn{7}{c}{$\rho$} \\
\cline{4-10}
 & & & -0.7 & -0.5 & -0.3 & 0 & 0.3 & 0.5 & 0.7 \\
\hline
MC &  20 & Q & 1.00 & 1.00 & 0.98 & 0.05 & 0.98 & 1.00 & 1.00 \\ 
  GC &  20 & Q & 0.99 & 0.97 & 0.83 & 0.04 & 0.96 & 1.00 & 1.00 \\ 
  APLE &  20 & Q & 1.00 & 1.00 & 0.98 & 0.05 & 0.98 & 1.00 & 1.00 \\ 
  RMC &  20 & Q & 1.00 & 1.00 & 1.00 & 0.06 & 1.00 & 1.00 & 1.00 \\ 
  RGC &  20 & Q & 0.13 & 0.11 & 0.09 & 0.04 & 1.00 & 1.00 & 1.00 \\ 
  RAPLE &  20 & Q & 1.00 & 1.00 & 1.00 & 0.06 & 1.00 & 1.00 & 1.00 \\ 
  GK &  20 & Q & 1.00 & 1.00 & 0.98 & 0.05 & 1.00 & 1.00 & 1.00 \\ 
  GK2 &  20 & Q & 0.98 & 0.85 & 0.52 & 0.04 & 1.00 & 1.00 & 1.00 \\ 
   \hline
\end{tabular}
\end{table}

\newpage
Table A.7: Empirical power function for Laplace distribution (n=400) and Queen (Q) W matrix
\begin{table}[ht]
\centering
\begin{tabular}{lrlrrrrrrr}
  \hline
\multirow{2}{*}{Measure} & \multirow{2}{*}{n} & \multirow{2}{*}{$W$} & \multicolumn{7}{c}{$\rho$} \\
\cline{4-10}
 & & & -0.7 & -0.5 & -0.3 & 0 & 0.3 & 0.5 & 0.7 \\
\hline
MC &  20 & Q & 1.00 & 1.00 & 0.92 & 0.05 & 0.95 & 1.00 & 1.00 \\ 
  GC &  20 & Q & 0.99 & 0.91 & 0.60 & 0.04 & 0.84 & 1.00 & 1.00 \\ 
  APLE &  20 & Q & 1.00 & 1.00 & 0.92 & 0.05 & 0.95 & 1.00 & 1.00 \\ 
  RMC &  20 & Q & 1.00 & 1.00 & 0.92 & 0.05 & 0.97 & 1.00 & 1.00 \\ 
  RGC &  20 & Q & 1.00 & 0.99 & 0.84 & 0.05 & 0.95 & 1.00 & 1.00 \\ 
  RAPLE &  20 & Q & 1.00 & 1.00 & 0.92 & 0.05 & 0.97 & 1.00 & 1.00 \\ 
  GK &  20 & Q & 1.00 & 0.99 & 0.78 & 0.05 & 0.87 & 1.00 & 1.00 \\ 
  GK2 &  20 & Q & 0.99 & 0.92 & 0.62 & 0.05 & 0.81 & 1.00 & 1.00 \\ 
   \hline
\end{tabular}
\end{table}

\newpage
Table A.8: Empirical power function for Mixture distribution (n=400) and Queen (Q) W matrix
\begin{table}[ht]
\centering
\begin{tabular}{lrlrrrrrrr}
  \hline
\multirow{2}{*}{Measure} & \multirow{2}{*}{n} & \multirow{2}{*}{$W$} & \multicolumn{7}{c}{$\rho$} \\
\cline{4-10}
 & & & -0.7 & -0.5 & -0.3 & 0 & 0.3 & 0.5 & 0.7 \\
\hline
MC &  20 & Q & 0.96 & 0.39 & 0.01 & 0.99 & 1.00 & 1.00 & 1.00 \\ 
  GC &  20 & Q & 0.00 & 0.00 & 0.00 & 1.00 & 1.00 & 1.00 & 1.00 \\ 
  APLE &  20 & Q & 0.96 & 0.37 & 0.01 & 0.98 & 1.00 & 1.00 & 1.00 \\ 
  RMC &  20 & Q & 0.98 & 0.73 & 0.20 & 0.83 & 1.00 & 1.00 & 1.00 \\ 
  RGC &  20 & Q & 0.64 & 0.11 & 0.00 & 1.00 & 1.00 & 1.00 & 1.00 \\ 
  RAPLE &  20 & Q & 0.98 & 0.73 & 0.20 & 0.82 & 1.00 & 1.00 & 1.00 \\ 
  GK &  20 & Q & 1.00 & 0.88 & 0.47 & 0.16 & 0.90 & 1.00 & 1.00 \\ 
  GK2 &  20 & Q & 0.98 & 0.80 & 0.44 & 0.11 & 0.81 & 1.00 & 1.00 \\ 
   \hline
\end{tabular}
\end{table}

\end{document}